# Number of Information and its Relation to the Cosmological Constant Resulting from Landauer's Principle

Ioannis Gkigkitzis[1] Ioannis Haranas[2] Samantha Kirk[1]

[1] Departments of Mathematics, East Carolina University
124 Austin Building, East Fifth Street, Greenville
NC 27858-4353, USA
E-mail: gkigkitzisi@ecu.edu

[2] Department of Physics and Astronomy, York University
4700 Keele Street, Toronto, Ontario, M3J 1P3, CANADA
E-mail: yiannis.haranas@gmail.com

**Abstract**

Using a recent published formula for the number of information $N$ that results from Landauer's principle we obtain an expression for the cosmological constant $\Lambda$. Next, assuming the universe as a system of mass $M$ satisfying Landauer's principle and eliminating its mass $M$ from the given expression for the number of information, we obtain a new expression that agrees with the one derived by Lloyd. Furthermore, we modify the generalized entropy relation and three equivalent entropy expressions are obtained. Finally, in two different universes the time rate of change of the entropy is calculated. In a flat universe the time rate of the entropy is time independent and depends on fundamental constants of physics.

**Keys words:** Landauer's principle, number of information, nats, density parameter, flat universe, Hubble's parameter.

## 1. Introduction

Any physical system can register information just from its existence. Systems that dynamically evolve in time not only transform but also process information. In relation to the laws of physics we say that these laws determine the amount of information that a given system can register (i.e. number of bits or nats) as well as the number of elementary logic operations that the given system can perform (i.e. numbers of operations). In his original paper Landauer (1988) also cited by Lloyd (2002), made the statement that information is physical but also that all information is registered and processed by physical systems. Physical systems can be described in terms of information, and information processing is related to the system description by physical laws. Landauer's principle is a physical principle pertaining to the lower theoretical limit of energy consumption of a computation. Landaurer postulated that any "logically irreversible manipulation of information, such as the erasure of a bit or the merging of two computation paths, must be accompanied by a corresponding entropy increase in non-information bearing degrees of freedom of the information processing apparatus or its environment." (Bennett 2003).

Landauer's principle asserts that there is a minimum possible amount of energy required to change one bit of information, known as the Landauer's limit, and it is equal to:

$$E_{min} = k_B T \ln 2 \qquad (1)$$

where $k_B = 1.38 \times 10^{-23}$ J/K is Boltzman's constant, and $T$ is the temperature of the circuit. Therefore Landauer's energy formula can be written in terms of the information number $N \ne 1$ in the following way:

$$E_{min} = N k_B T \ln 2 \qquad (2)$$

Recent experimental studies provide further evidence that Landauer's prediction is true. For example, the equivalence between information and energy can be interpreted using the results obtained by a recent experiment by Funo, et al., (2012). In their experiments the authors have shown that entanglement can produce an increase or gain of thermodynamic work, where the gain is determined by the change of the information content. Similarly, Berut et al., (2012) have shown that there is a link between information theory and thermodynamics. Landauer's principle is a simple consequence where its logic emanates directly from the second law of thermodynamics. The law states that the entropy of a closed system cannot decrease together with the corresponding temperature. Therefore, if one nat of information is lost during a computation, the amount of entropy generated is at least $k_B \ln 2$, and therefore the energy emitted in the environment is $E \ge k_B T \ln 2$. Landauer's principle has been accepted as a physical law, but it has also been challenged by Shenker (2000) and Norton (2004) and defended by Bennett (2003) and Ladyman et al. (2008).

In relation to the number of information $N$ in cosmology in two recent papers by Haranas and Gkigkitzis (2013a, 2013b), the authors have investigated the Bekestein bound of information number $N$ and its relation to cosmological parameters in a universe with and without cosmological constant. In this paper the authors have derived the dependence of Hubble's parameter on the number of information $N$ (Haranas and Gkigkitzis 2013a). Furthermore, in a recent paper by Alfonso Faus and Fullana i Alfonso (2013) the authors interestingly propose a cosmic background, described by lower energy quantum state particles of a Bose condensation that is present anywhere. Furthermore, the authors identify the quantum of the self gravitational potential energy of any particle, with the bit of information of a minimum energy. In our effort to investigate the relation of information and cosmology, we will use the expression for the entropy given by Alfonso Faus and Fullana i Alfonso (2013).

## 2. Information and its relation to the cosmological constant

With reference to Alfonso Faus and Fullana i Alfonso (2013) the authors say that "all physical systems of mass $M$ and energy $Mc^2$ are equivalent to an amount of information in number of bits" of the order of:

$$N \cong \left( \frac{Mc^2}{\hbar H} \right), \qquad (3)$$

where $M$ is the mass of the system, $c$ is the speed of light, $\hbar$ is Planck's constant, and $H$ is the Hubble constant. In Alfonso Faus and Fullana i Alfonso (2013) the authors claim that Eq. (3) is of "universal validity". They further claim that that the unit of energy, should be taken as the minimum quantum of energy $\hbar H$. This will also imply that the relativistic energy of a mass $M$ is equal to $N$ times this minimum quantum of energy, $N\hbar H$ where $N$ is the number of information in bits or nats. In other words the product $N\hbar H$ corresponds to the energy of all the information number $N$ that a particular system carries. Using the expression for the entropy in terms of the number of information $N$ in nats as it is given in Haranas and Gkigkitzis (2013a) and replacing $N$ by Eq. (2), Alfonso Faus and Fullana i Alfonso (2013) obtain a generalized relation for the entropy that is given by:

$$S = N k_B \ln 2 = \left(\frac{Mc^2}{\hbar H}\right) k_B \ln 2. \tag{4}$$

In the case that $N = \left(\dfrac{Mc^2}{\hbar H}\right) = 1$ that would imply that $Mc^2 = \hbar H$ and therefore the entropy associated with one nat is equal to:

$$S = k_B \ln 2. \tag{5}$$

In our effort to establish the validity of Alfonso Faus' and Fullana's generalized relation of entropy, let as investigate if the relation can lead us to any results known in today's cosmology. We are interested in investigating if Eq. (4) can lead us to a known expression for the cosmological constant $\Lambda$. With reference to Haranas and Gkigkitzis (2013a) we say that the entropy of a universe that involves a cosmological constant can be written as:

$$S = \frac{3\pi k_B}{\Lambda \ell_p^2}, \tag{6}$$

where $\ell_p^2 = G\hbar/c^3$ is the Planck length. From Eqs (4) and (6) we obtain that the cosmological constant can be written as:

$$\Lambda = \frac{3\pi}{\ln 2}\left(\frac{\hbar H}{Mc^2 \ell_p^2}\right), \tag{7}$$

where $M$ is the mass of the universe that is assumed to be a closed system. Eliminating the Planck length we obtain:

$$\Lambda = \frac{3\pi}{\ln 2}\left(\frac{Hc}{MG}\right), \tag{8}$$

furthermore, we can eliminate the mass of the universe using one of the equations derived in Haranas and Gkigkitzis (2013a), which is in agreement with Hoyle (1958) and Valev (2002) namely:

$$M = \frac{c^3}{GH}, \qquad (9)$$

substituting Eq. (9) in (8) we obtain that the cosmological constant can be finally written as:

$$\Lambda = \frac{3\pi}{\ln 2}\left(\frac{H}{c}\right)^2 = 13.60\left(\frac{H}{c}\right)^2. \qquad (10)$$

Equation (10) is basically identical with that given in Islam (1999):

$$|\Lambda| = 21\left(\frac{H}{c}\right)^2. \qquad (11)$$

Therefore, we have derived a known expression for the cosmological constant considering the universe to be a system of mass *M*, and accepting that there is a number of information *N* in nats associated with its mass. This is achieved after deriving a new expression for the cosmological constant that itself involves basic fundamental physics constants, i.e. Eq. (8), from which we eliminate the mass of the universe using an expression derived in Haranas and Gkigkitzis (2013a). The expression obtained for the cosmological constant involves a numerical coefficient that is 1.54 times the coefficient that is given in by Islam (1999). This derivation constitutes a check that introduces merit to the definition of the number of information *N* as it is given in Eq. (3).

## 3. Discussion and results

We have considered Landauer's principle in an effort to investigate the idea of information and its relation to physical systems. We have used the equation for the definition of the number of information *N*, as it is given by Alfonso Faus and Fullana i Alfonso (2013) in Eq. (3). Considering a universe of mass *M*, and using a derived expression that relates the entropy *S* to the cosmological constant *Λ*, we obtain an intermediate expression for the cosmological constant. Next, eliminating the mass of the universe we obtain a known expression for the cosmological constant as it is given in Islam (1999). Furthermore, taking the whole universe as a closed system we derive an alternate expression for the cosmological constant that now depends on fundamental constants of physics namely the Hubble constant *H*, the speed of light *c*, the mass of the universe *M*, and the gravitational constant *G*, i.e. Eq. (8). With reference to the definition of the information bit number *N* as given by Alfonso Faus and Fullana i Alfonso (2013) in the case where the universe is considered to be a closed system of mass *M*, and eliminating the mass *M* using Eq. (9) we obtained that:

$$N = \frac{1}{(Ht_p)^2} = \left(\frac{t}{t_p}\right)^2, \qquad (12)$$

where *t* is the age of the universe at any time. Our result is in agreement with Lloyd (2002) where the author predicts that this is equal to the maximum number of bits registered by the universe using matter, energy, and gravity, and it is found with the help of the Bekenstein bound and the holographic principle of the universe as a whole. It is given by the square of the ratio of the age of the universe to that of Planck time. Equation (12) demonstrates that at $t = t_p$ the number of information $N = 1$. In other words when $t = t_p$ there exists at least on nat of information in the universe that is possible to be decompressed through matter and energy. Next, using the equation of the Hubble constant that depends on the information number $N$ as it is given in Haranas and Gkigkitzis (2013b) namely:

$$H = \left(\frac{\pi}{\ln 2}\right)^{1/2} \left(\frac{c}{\ell_P \sqrt{N}}\right) = \left(\frac{\pi}{\ln 2}\right)^{1/2} \frac{1}{t_P \sqrt{N}}, \tag{13}$$

where $\ell_p$ and $t_P$ are the Planck and length and time, and substituting in Eq. (3) we obtain a modified expression for the number of information in nats, and therefore Eq. (3) becomes:

$$N \cong \frac{\xi_0 M c \ell_P}{\hbar} \sqrt{N} \cong \xi_0 \left(\frac{\ell_P}{\lambda_{Co_U}}\right) \sqrt{N}, \tag{14}$$

where $\lambda_{C_U}$ is the Compton wavelength corresponding to the universe mass $M$, and $\xi_0 = (\ln 2/\pi)^{1/2}$. Solving Eq. (14) for $N$ we obtain that:

$$N = N_1 = 0 \tag{15}$$

$$N = N_2 \cong \left(\xi_0 \ell_P \frac{Mc}{\hbar}\right)^2 \cong \xi_0^2 \left(\frac{\ell_P}{\lambda_{C_U}}\right)^2 = O(10^{120}). \tag{16}$$

Using the second part on the RHS of Eq. (13) and substituting $\ell_p^2 = G\hbar/c^3$ Eqs. (15) and (16) can be written as:

$$N = N_1 = 0$$

$$N = N_2 \cong \frac{\ln 2}{\pi} \left(\frac{M_U}{m_p}\right)^2 = O(10^{120}). \tag{17}$$

Therefore the entropy of the universe can be written in the following alternative ways that involve fundamental parameters of physics namely:

$$S = Nk_B \ln 2 = \left(\frac{Mc^2}{\hbar H}\right) k_B \ln 2 = \frac{k_B}{H^2 t_P^2} \ln 2 = \left(\frac{t}{t_P}\right)^2 k_B \ln 2 = \frac{\ln 2}{\pi} \left(\frac{M_U}{m_P}\right)^2 k_B \ln 2. \tag{18}$$

The present value of the entropy at the horizon of the universe can be found by using the relation of the horizon radius $R_H = c/H_0$ and therefore from Eq. (18) we obtain that:

$$S_0 = Nk_B \ln 2 = \left(\frac{Mc}{\hbar} R_H\right) k_B \ln 2 = \left(\frac{R_H}{\lambda_{C_U}}\right) k_B \ln 2 = \ln 2 \frac{k_B}{t_P^2} \left(\frac{R_H}{\ell_P}\right)^2, \tag{19}$$

Eq. (19) demonstrates that the entropy of the horizon increases as the universe expands and $R_H$ increases. Furthermore, with reference to Haranas and Gkigkitzis (2013b) and using the derived equation that relates

Hubble's constant to the number of information *N*, i.e. Eq (13), and assuming that the information changes with time *t* we write the entropy of the universe as a function of time and from that we calculate its time rate of change and therefore we have:

$$S(t) = \left(\frac{M_U c^2}{\hbar H(t)}\right) k_B \ln 2. \tag{20}$$

Upon substituting Eq. (13) into (20) and simplifying we have that:

$$S(t) = \frac{\ln 2 \sqrt{\ln 2}}{\sqrt{\pi}} \left(\frac{M_U c}{\hbar}\right) \ell_P k_B \sqrt{N(t)}, \tag{21}$$

taking the derivative with respect to time *t* and simplifying we obtain that:

$$\frac{dS(t)}{dt} = \frac{\ln 2 \sqrt{\ln 2}}{2\sqrt{\pi}} \left(\frac{M_U c}{\hbar}\right) \ell_P k_B \frac{1}{\sqrt{N(t)}} \frac{dN(t)}{dt}. \tag{22}$$

In a flat universe (i.e. curvature constant $k = 0$) and density parameter $\Omega = 1$ we find that the information number as a function of time is given by the following expression (Haranas and Gkigkitzis, 2013):

$$N(t) = \frac{4\pi}{\ln 2}\left(\frac{c}{t_P}\right)^2 t^2, \tag{23}$$

and therefore we can easily prove by substituting for *N(t)* using Eq. (23) together with its time derivative that the quantity $(N(t))^{-1/2} dN(t)/dt$ takes the value:

$$\frac{1}{\sqrt{N(t)}} \frac{dN(t)}{dt} = \frac{8\pi \sqrt{\ln 2}}{2 \ln 2 \sqrt{\pi}} \left(\frac{c}{\ell_P}\right). \tag{24}$$

Next, substituting Eq. (24) in (22) we find that the rate of change of the entropy of the universe as a function of time becomes:

$$\frac{dS(t)}{dt} = 4 \ln 2 \left(\frac{Mc^2}{\hbar}\right) k_B = 4 \ln 2 \left(\frac{c}{\lambda_{C_U}}\right) k_B, \tag{25}$$

where $\lambda_{C_U} = \hbar / M_U c$ is the Compton wavelength that corresponds the mass of the universe $M_U$, and thus the time rate of change of the universe entropy takes the numerical value:

$$\frac{dS(t)}{dt} = 4.676 \times 10^{81} \text{ J K}^{-1} \text{ s}^{-1}. \tag{26}$$

For a flat universe the rate of change of the entropy is constant, and depends only on fundamental parameters of physics. The units of J K$^{-1}$s$^{-1}$ is equivalent to is power per degree Kelvin or Watt K$^{-1}$. Using the age of the universe given by Bennet et al. (2013) namely 13.798 by = $4.346 \times 10^{17}$ s, we obtain that the total entropy of the universe is:

$$S = 2.032 \times 10^{99} \text{ J K}^{-1}. \tag{27}$$

This is in agreement with the order of magnitude entropy budget given in Table 2 of Egan, and Lineweaver (2010), where the authors give that $S = (2.6 \pm 0.3) \times 10^{122} k_B = (3.588 \pm 0.3) \times 10^{99}$ J K$^{-1}$. Similarly in the case of a closed universe i.e. curvature constant $k = 1$ and density parameter $\Omega > 1$ we obtain that the information number as a function of time is given by the following expression (Haranas and Gkigkitzis, 2013):

$$N(t) = 6(2 + \Omega_1) A^2 \tan^2\left(t\sqrt{\frac{2+\Omega_1}{6}}\right) \qquad (28)$$

and therefore the quantity $(N(t))^{-1/2} dN(t)/dt$ takes the value:

$$\frac{1}{\sqrt{N(t)}} \frac{dN(t)}{dt} = \frac{2\sqrt{\pi}(2+\Omega_1)}{t_p \sqrt{\ln 2}} \sec^2\left(t\sqrt{\frac{2+\Omega_1}{6}}\right), \qquad (29)$$

where $1 - \Omega = -\Omega_1 < 0$ and therefore the rate of change of the entropy becomes:

$$\frac{dS(t)}{dt} = 2\ln 2 (2+\Omega_1)\left(\frac{M_U c^2}{\hbar}\right) k_B \sec^2\left(t\sqrt{\frac{2+\Omega_1}{6}}\right), \qquad (30)$$

Integrating Eq. (31) with respect to $t$ we obtain

$$S(t) = 2\sqrt{6}\ln 2 \left(\frac{M_U c^2}{\hbar}\right) k_B (2+\Omega_1) \tan\left(t\sqrt{\frac{2+\Omega_1}{6}}\right)\bigg|_0^{t=t_{age}=4.346 \times 10^{17}}. \qquad (31)$$

Similarly, with $\Omega_1 = 0.2$ obtain that:

$$S = 2.240 \times 10^{83} \text{ J K}^{-1}.$$

Finally, assuming that $M$, $\hbar$, are constant and since $H$ increase with time, a consequence of Eq. (4) is that a variable speed can be possibility, without violating the second law of thermodynamics.

## 4. Conclusions

Using a generalized relation for the entropy that involves a new definition of the number of information $N$, we have obtained an expression of the cosmological constant $\Lambda$ that involves only fundamental parameters of physics. Thus, by eliminating the mass of the universe we obtain the known expression for the cosmological constant as it is given by Islam. In our opinion this result constitutes a check on the merit of the definition of the number of information $N$ according to Alfonso Faus and Fullana i Alfonso (2013). Moreover, taking the whole universe as a closed system we derive an alternate expression for Faus' and Fullana's number of information $N$, as the square of the ratio of the age of the universe at any time over the Planck time, a result that is agreement with Lloyd (2002). This can also be expressed as the square of the ratio of the mass of the universe over the Planck mass. We also find that the present era value of the entropy at the horizon of the universe is proportional to the square of the horizon radius because the number of information $N$ itself is proportional to the horizon radius itself. Finally, we have written the entropy of the universe as a function of a time depended number of information *N* via which the time rate of entropy for two different kinds of universes has been calculated, and from that the entropy as a function of time has also been derived and numerical value estimates have been obtained.

**Acknowledgements**
The authors want to thank an unknown reviewer who with his/her valuable comments help improved this manuscript.